\documentclass[doublecol]{epl2} 
\usepackage{epsfig}
\DeclareGraphicsExtensions{.eps, .jpg}
\input epsf

\title{Pressure-temperature phase diagram of  charge ordering in Nd$_{1/2}$Sr$_{1/2}$MnO$_3$} 
\author{P. Maselli\inst{1}  \and D. Nicoletti\inst{1}  \and A. Nucara\inst{1}  \and F. M. Vitucci\inst{1}  \and A. Irizawa\inst{2}  \and K. Shoji\inst{2}  \and T. Nanba\inst{2} \and P. Calvani\inst{1}}
\shortauthor{P. Maselli \etal}

\institute{                    
  \inst{1} CNR-SPIN and Dipartimento di Fisica, Universit\`a di Roma La Sapienza, Roma, Italy\\
  \inst{2} Graduate School of Science and Technology, Kobe University, Hyogo, Japan\\}

\abstract{
We observe how the charge-ordering temperature $T_{CO}$ of  Nd$_{1/2}$Sr$_{1/2}$MnO$_3$ decreases with the external pressure $p$ from 160 K at $p =0$ down to 30 K at $p \simeq$ 4.5 GPa, by measuring the values $p,T$ where the far-infrared spectral weight of the metallic phase is fully recovered. We thus determine the ($p,T$)  phase diagram of CO in that manganite. We also find that the parameter $d(ln T_{CO})/dp$  which describes this metallization from the CO phase is equal and opposite to the quantity  $d(ln T_{c})/dp$ which governs the metallization of the paramagnetic state at comparable Curie temperatures $T_c$, in similar manganites at half doping. 
}

\begin{document}

\maketitle

\section{Introduction}
The $x,T$ phase diagram of the  manganites  A$_{1-x}$B$_{x}$MnO$_3$, where A is a  lanthanoid and B  an alkali earth,  includes charge-ordered phases for   $0.5 \leq x \leq 0.85$  and  $T < T_{CO}$.  $T_{CO}$,  the charge-ordering (CO) temperature,  in the absence of an external field may vary between  160 K for  Nd$_{0.5}$Sr$_{0.5}$MnO$_3$ (NSMO)  \cite{Caignaert} to more than 500 K for Bi$_{0.5}$Sr$_{0.5}$MnO$_3$  \cite{Frontera}. The CO  is usually associated with both orbital  ordering and antiferromagnetism (AFM)  \cite{Dagotto}, so that the ordered  phase may collapse under magnetic fields $B$ which span from 10 T for NSMO \cite{Popov} to more than 50 T for Bi$_{0.5}$Sr$_{0.5}$MnO$_3$ \cite{Frontera}. For a long time,  CO in manganites has been  discussed in terms of arrays of Jahn-Teller small polarons \cite{Good}.  However, in manganites with $T_{CO}$ below room temperature and with moderate critical fields, there are strong indications  \cite{Herrero,Loudon1,Nucara,Milward}  that the CO phase is better described in terms of Charge Density Waves (CDW) \cite{Gruner}.  This instability of the Fermi liquid takes place in a regime of moderately weak electron-phonon coupling and is associated with a perfect (in the ideal one-dimensional CDW)  or partial nesting of the Fermi surface. 

In the optical conductivity  $\sigma (\omega)$, the transition to a CO   phase  shows up in the appearance of an insulating gap $E_g$ in the mid infrared \cite{Calvani98} and of absorption peaks in the Terahertz range \cite{Nucara,Zhukova}. Conversely,  the infrared-gap filling- with partial shielding of the  phonon peaks by a free-carrier absorption - provides evidence for an insulator-to-metal transition (IMT) due to the CO collapse.    
 Besides $T$ and $B$, an external pressure $p$ \cite{Sacchetti07} can destabilize the charge order. Pressures in the GPa range have been shown to destroy the CO ground state of CeTe$_3$ at room temperature \cite{Sacchetti07}.  In the manganites, no  infrared measurements at high pressure  have been reported in a CO state to our knowledge. However,  transitions from the paramagnetic (PM) phase to the ferromagnetic (FM), metallic phase,  induced by pressure,  were observed by monitoring either  the mid  \cite{Congeduti} or the far infrared \cite{Sacchetti06} $\sigma (\omega)$. In the present experiment we have investigated the effect of $p$ on  the CO phase of a manganite at low temperature,  by measuring its $\sigma (\omega)$ in the far infrared at different pressures and temperatures. 
The selected system is Nd$_{1/2}$Sr$_{1/2}$MnO$_3$ (NSMO), a manganite which exhibits both a homogeneous, commensurate CO with periodicity 2$a$ ($a$ = lattice parameter), and  AFM (of charge-exchange, CE, type) below a  $T_{CDW}\simeq$ 160 K (onset) \cite{Caignaert} at ambient pressure.  This transition is clearly detected in the infrared through a CO gap at $E_g \sim$ 0.1 eV \cite{Nucara}, which opens in the Drude term of  the ferromagnetic (FM) metallic phase. This is stabilized by  the double-exchange mechanism  \cite{Dagotto} from  $T_{CO}$ to a Curie temperature $T_c \simeq$  255 K \cite{Tokura}. Here we detect the IMT induced by pressure at   $T < T_{CO}$ by  measuring  the spectral weight 

\begin{equation}
W(p,T)=\int_{\Omega_1} ^{\Omega_2} \sigma(\omega,T)\mathrm{d}\omega
\label{W}
\end{equation}

\noindent
at zero pressure in the metal at 200 K and observing how it is recovered in the low-$T$ AFM phase due to closure of the CO gap. In our case, $\Omega_1$ = 120 cm$^{-1}$ and $\Omega_2 =$ 1000 cm$^{-1}$.

\section{Experimental}
The optical conductivity of NSMO was extracted from reflectivity measurements  performed down to 30 K in a diamond anvil cell (DAC). In order to obtain a reasonable photon flux through the small aperture of the  DAC  \cite{Cestelli}, we used instead of conventional sources the InfraRed Synchrotron Radiation (IRSR)   available at the beam line BL43IR of the storage ring SPring-8 \cite{Ikemoto}.  

The sample, with a reflecting  surface of about 100 $ \times$ 100 $\mu$m corresponding to the $ab$ plane, was cut from a single crystal grown by the floating-zone method \cite{Herrero}. It was  inserted in a DAC, together with a parallel gold  mirror for reference, and with a small ruby. The frequency of its fluorescence peak provided at any $T$, after long thermalization, the pressure $p$ at $T$. The pressure medium was Daphne7474 \cite{Murata} and the cell was inserted in a liquid helium cryostat. Due to the  surprising low pressures at which the AFM-FM transition occurs, the contraction of the cell upon cooling played a major role. For this reason, one had often to select different  $p$ values for the measurements at different  $T$. 


\begin{figure}[!t]   \begin{center}  
\leavevmode    
\epsfxsize=8cm \epsfbox {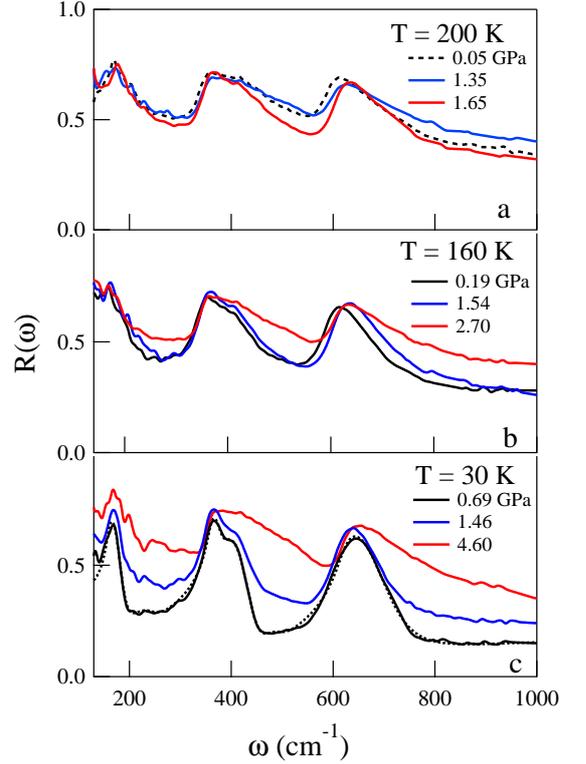}    
\caption{(Color online). Reflectivity $R(\omega)$ of Nd$_{1/2}$Sr$_{1/2}$MnO$_3$  at three different temperatures and under increasing pressures, as measured at the interface with the diamond window. The dotted line in (c) is an example of Drude-Lorentz fit to data, as explained in the text.}
\label{R-sd}
\end{center}
\end{figure}


The  intensity $I_s$ reflected by the sample was divided by the $I_r$ reflected in the same conditions by the mirror, to obtain the reflectivity $R(\omega)$ at the interface between the manganite and the diamond window.  This quantity is shown in Fig. \ref{R-sd} at different temperatures and pressures. 
Even these raw data clearly show the effect of pressure on the optical response. In a) at 200 K (in the FM phase) $R(\omega)$ increases slightly at high $\omega$ and then \textit{decreases} in the whole range for increasing $p$. In c) at 30K, deeply in the CO ground state, both the overall reflectivity  and the phonon shielding  \textit{increase} with pressure. At 4.6 GPa the  shape of $R(\omega)$ is again similar to that observed in panel a) at $p \sim 0$. At 160 K (b), the metallic $R(\omega)$ is reached at 2.7 GPa.  Due to the limited frequency range accessible to the experiment, standard Kramers-Kronig transformations could not be used to extract $\sigma (\omega)$  from $R(\omega)$. This quantity  was instead obtained by fitting, to the data of Fig. \ref{R-sd}, the formula

\begin{equation}
R(\omega) = {\left | \frac{\sqrt {\tilde \epsilon (\omega)} - n_d}{\sqrt {\tilde \epsilon (\omega)} + n_d} \right |}^2
\label{fit}
\end{equation}

\noindent
where the refractive index of diamond $n_d$ was kept real and equal to 2.37, a value obtained by averaging $n_d (\omega)$  \cite{Dore} over the whole measuring range. An example of such a fit is shown in Fig. \ref{R-sd}-c.


\begin{figure}[!t]   \begin{center}  
\leavevmode    
\epsfxsize=8cm \epsfbox {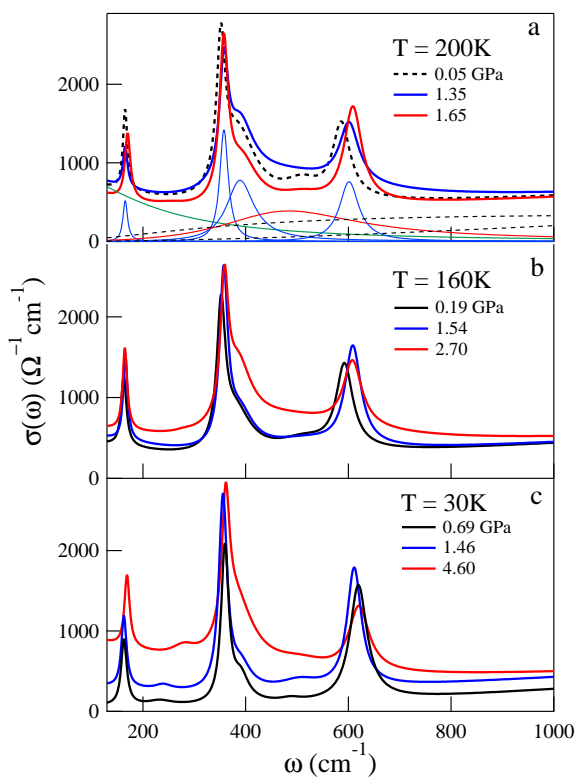}    
\caption{(Color online). Optical conductivity $\sigma(\omega)$ of Nd$_{1/2}$Sr$_{1/2}$MnO$_3$, as extracted from the $R(\omega)$  at three different temperatures under increasing pressures.  Individual phonon contributions to the fit  (thin lines), a polaron-like band centered around 500 cm$^{-1}$, and tails of high-frequency bands (dashed lines) are also shown in panel a).}
\label{sigma}
\end{center}
\end{figure}


For the dielectric function $\widetilde\epsilon(\omega)$ we assumed the usual expansion in terms of Drude and Lorentz  oscillators from  free charges and  phonons, respectively. The fit also requires a  polaron-like excitation \cite{Calvani} centered around 500 cm$^{-1}$, and the inclusion of the tails of the higher-frequency bands previously measured in NSMO \cite{Nucara08}. An example of such decomposition is shown at 200 K in panel a) of Fig. \ref{sigma}, where $\sigma(\omega)= \omega \epsilon_2(\omega)/4\pi$ is reported for the same temperatures and pressures as in Fig. \ref{R-sd}. In the FM phase at 200 K (Fig.  \ref{sigma}-a) one sees again that at the highest  $p$ the sample has  partly lost its  metallic character. In the CO-AFM phase (panels b) and c)), the effect of pressure is opposite. It promotes a metallization process which shows up clearly both at 30 K (Fig. \ref{sigma}-b) and at 160 K (c), through an increase of the overall conductivity and a partial shielding of the phonon peaks. 

\section{Results and discussion}
\begin{figure}[!t]   \begin{center}  
\leavevmode    
\epsfxsize=8cm \epsfbox {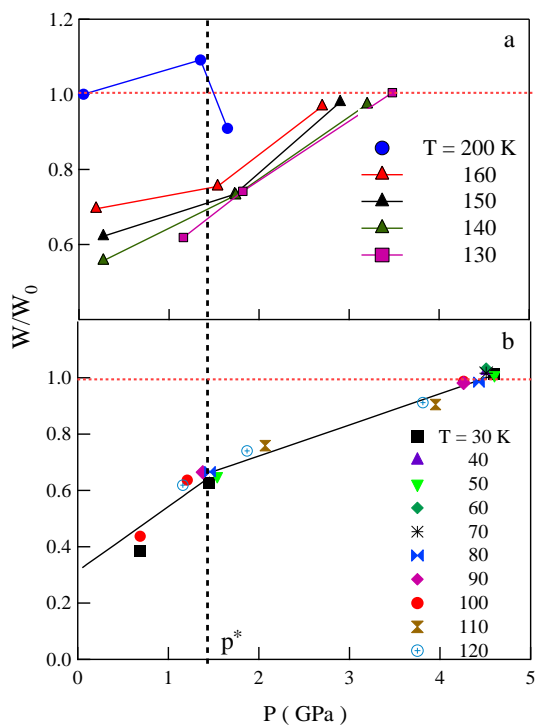}    
\caption{(Color online). Spectral weight vs. $p$ and $T$ of Nd$_{1/2}$Sr$_{1/2}$MnO$_3$, as calculated from Eq. \ref{W} and normalized to the value $W_0$ in the metallic phase at 200 K and $p$ = 0.05 GPa. The lines are guides to the eye.}
\label{weight}
\end{center}
\end{figure}

In order to describe quantitatively  the metallization process, we have calculated $W$ and normalized its value to the corresponding $W_0$ in the metallic FM phase at the lowest pressure (T=200 K, P=0.05 GPa). The results  are reported as a function of pressure in Fig. \ref{weight}. For sake of clarity, 
data taken above 130 K  are shown in panel a, those below this temperature in panel b.  In the FM  phase at $T=200 K$ (Fig. \ref{weight}-a), the behavior of $W$  indicates  a significant decrease in the metallic character of NSMO around $p^* \simeq$ 1.5 GPa. This effect,  similar to that reported in La$_{0.75}$Ca$_{0.25}$MnO$_3$ \cite{Ding}, points out the increase of  the super-exchange interaction, which tends to establish an AFM coupling between Mn nearest neighbors. In Fig. \ref{weight}-a, the drop in $W$ occurs at a much lower pressure than in La$_{0.75}$Ca$_{0.25}$MnO$_3$ (23 Gpa), but in this compound the ground state is FM, while in NSMO the FM phase is unstable with respect to the AFM ground state. The decrease of $W$ at 200 K under pressure may then be attributed to  the appearance of AFM zones, in a phase-separation scenario often invoked for the manganites at half doping \cite{Dagotto}. 

In the four sets of data  at or below 160 K in Fig. \ref{weight}-a,  $W$ increases monotonically with $p$. In the end, the metallic spectral weight $W_0$ is recovered at  a pressure $p_M$ which increases for decreasing temperature. One thus can associate with each $p_M$ a $T_{CO}$  which marks the collapse of the CO  phase. Those pairs of values will be used below to plot a  boundary line between the CO and the metallic phase of NSMO.
Below 130 K  (Fig. \ref{weight}-b),  all data show a similar behavior and therefore they are grouped around a single line. At the lowest $T$ $W \simeq W_0$ at $p_M \simeq$ 4.5 GPa, which therefore is the CO critical pressure. This value is in excellent agreement with the dc resistivity which, above 4.5 GPa, does not show any CO transition  \cite{Yu} and is rather low. This  confirms that  the interactions which stabilize the CO-AFM phase in manganites like NSMO are not so strong as to prevent the application of the CDW model \cite{Nucara}. 


Let us now discuss what is implied in the above results.  Pressure promotes the IMT by reducing either  the average Mn-O distance $d$ and  the  average  angle $\beta$  of bonds Mn-O-Mn in the $ab$ plane.Then,  $\pi - \beta$ measures the distortion of the $MnO_6$ octahedra \cite{Laukhin}. Both mechanisms affect directly the spectral weight $W$ through the electron bandwidth, and in a tight binding approximation  \cite{Radaelli} 

\begin{equation}
W(p,T<T_{CO})\approx\frac{\cos[(\pi - \beta)/2]}{d^{3.5}}							
\label{fit}
\end{equation}

\noindent
However, $d$  decreases by a bare 0.7 \% between 0 and 4.5 GPa, while in the ground CE state $\beta$ =  157$^0$ \cite{Kajimoto}. Even assuming $\beta = \pi$ at  $p_M$,   according to Eq. \ref{fit} the isotropic octahedra increase  $W$ by just a 4 \%. In Fig. \ref{weight}-b, $W$ grows instead by a factor of 3. Therefore, the small  pressure-induced changes in Eq. \ref{fit} are hugely amplified  by non-linear effects. The most important is obviously the transition from the AFM superexchange to the FM double-exchange between the Mn$^{3+}$ and Mn$^{4+}$ ions. However, also the charge-lattice  interaction is expected to play a role via the Jahn-Teller effect \cite{Laukhin}.  

The change of slope at $p^*$ in Fig. \ref{weight}-b may be related to an observation  \cite{Arulraj}  in the similar compound  Nd$_{1/2}$Ca$_{1/2}$MnO$_3$. Therein, for increasing $p$, while the lattice parameter $a$ decreases monotonically,  the Jahn-Teller distortion was found  to decrease up to 7 GPa, and then to increase again.  Its "V"-shaped behavior  vs. $p$ was attributed to a subtle shear strain of the octahedra. In NSMO this should happen at   pressures lower than 7 GPa,  due to the larger ionic volume of Sr with respect to Ca, and may explain the observation at $p^*$. However, the origin of the crossover in  Fig. \ref{weight} at the same $p^*$ in both  phases FM and CO, is not clear yet.


\begin{figure}[!t]   \begin{center}  
\leavevmode    
\epsfxsize=8cm \epsfbox {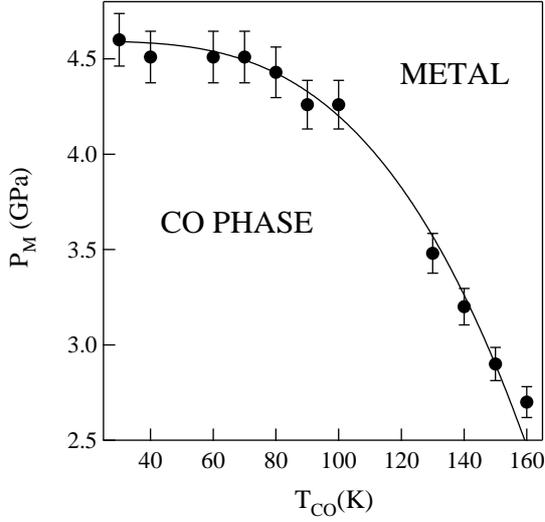}    
\caption{(Color online). Pressure-temperature  diagram of the CO phase in Nd$_{1/2}$Sr$_{1/2}$MnO$_3$.  $p_M$ and $T_{CO}$ are  the pressure and the temperature, respectively, where $W  \simeq W_0$ in Fig. \ref{weight} (dots). The phase boundary is a power-law fit to data.} 
\label{diagram}
\end{center}
\end{figure}

\begin{figure}[!t]   \begin{center}  
\leavevmode    
\epsfxsize=8cm \epsfbox {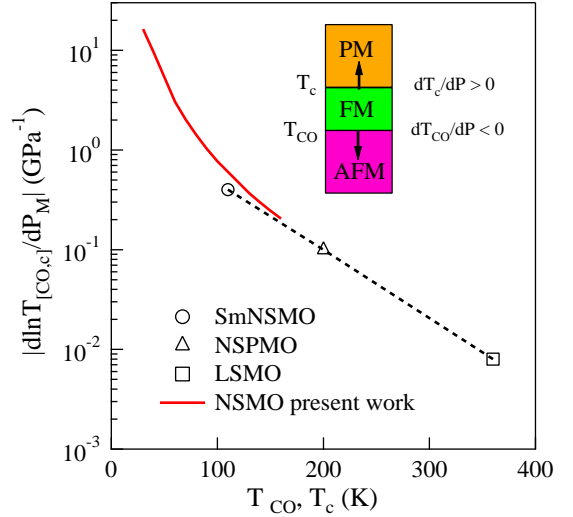}    
\caption{(Color online).  Absolute  values of the $\Psi$ defined in Eq. \ref{Psi-eq}. Those  calculated from the  metal-CO boundary line (solid line) are negative, those  measured at $T_c$ (PM-FM transition)in other manganites at half doping are positive. The latter ones (empty symbols, \cite{Zhao}) refer to : SmNSMO = (Sm$_{0.875}$Nd$_{0.125})_{0.5}$Sr$_{0.5}$MnO$_3$ ($T_c$ = 115 K); NSPMO = Nd$_{0.5}$Sr$_{0.36}$Pb$_{0.14}$MnO$_3$ ($T_c$ = 204 K);  LSMO = La$_{0.5}$Sr$_{0.5}$MnO$_3$ ($T_c$ = 360 K). The dashed line is a guide to the eye. The scheme on the right illustrates the processes which, either starting from the PM or from the CO-AFM phase, expand the FM metallic phase. They in fact lead to  positive and negative values of $\Psi$, respectively.}
\label{Psi}
\end{center}
\end{figure}

The pairs of values [$p_M (T),T_{CO}$]  obtained  from Fig. \ref{weight} at $W /W_0 \simeq 1$ allow one to build up the  boundary line between the metallic and the CO phase reported in Fig. \ref{diagram}, which is obtained by a power-law fit to data (dots). The transition pressure increases steeply below $T_{CO}$, to saturate at about 4.5 GPa at low temperature. This behavior is well illustrated by the variation of the parameter \cite{Font, Zhao}  

\begin{equation}
\Psi =  (1/T_M) dT_{M}/dp_M = d(lnT_{M})/dp_M
\label{Psi-eq}
\end{equation}

\noindent
which was originally reported for the PM-FM transition at the Curie temperature $T_c$ and which here can be calculated from the derivative of the boundary line in Fig. \ref{diagram} (see the scheme in the inset of Fig. \ref{Psi}). Therefore, in Eq. \ref{Psi-eq}, $T_M = T_c$  for the PM-FM metallization ($\Psi > 0$) and $T_M  = T_{CO}$  for the  AFM-FM transition ($\Psi < 0$).  In Fig. \ref{Psi}  $\Psi$ is plotted as extracted from the present data (red solid line) and compared with the (dashed) line reported in Ref. \cite{Zhao}. This was traced through three values (empty symbols), having opposite sign with respect to the present ones,  and taken around  $T_c$  in other manganites at half doping: SmNSMO = (Sm$_{0.875}$Nd$_{0.125})_{0.5}$Sr$_{0.5}$MnO$_3$ ($T_c$ = 115 K); NSPMO = Nd$_{0.5}$Sr$_{0.36}$Pb$_{0.14}$MnO$_3$ ($T_c$ = 204 K);  LSMO = La$_{0.5}$Sr$_{0.5}$MnO$_3$ ($T_c$ = 360 K) . The agreement both in the slope  and in the absolute values of $\Psi$ in the common range of transition temperatures is impressive, but can be explained by the following argument. Indeed \cite{Font},  $T_c \propto D$ and $T_{CO}  \propto U/D$ where $D$ is the carrier bandwidth  renormalized by  polaronic effects \cite{Millis} and $U$ is the correlation energy (assumed to be independent of $p$). Therefore, at the same $T$, $d(lnT_{c})/dp_M = -d(lnT_{CO})/dp_M$.  


\section{Conclusion}
We have  reported here the optical observation of the collapse, induced by pressure, of the charge-ordered phase  in NSMO at low temperature. The process  slows down at about the same pressure $p^*$ where the metal shows an instability above  $T_{CO}$. Through the analysis of the spectral weight we obtain the $p-T$  phase diagram of the charge order and the parameter $\Psi$ of the AFM-FM transition. This is equal and opposite to that measured for the PM-FM transition at comparable Curie temperatures  in other manganites at half doping. This confirms experimentally that  the metallization  induced by pressure in manganites is governed by the same physics, either if one starts from a paramagnetic insulator, or from a charge-ordered antiferromagnet.

\acknowledgments
We thank X. Blasco for providing  the NSMO crystal  which the present sample was cut from.
 


\end{document}